\documentclass[prl,aps,twocolumn,twoside,showkeys]{revtex4-1}

\usepackage{amsmath,amsfonts,amssymb,amsthm,mathtools}

\usepackage[english]{babel}
\usepackage[utf8]{inputenc}
\usepackage[T1]{fontenc}

\usepackage{latexsym}
\usepackage{graphicx}
\usepackage{epstopdf}

\usepackage{subfigure}
\usepackage{MnSymbol}
\usepackage{hyperref}
\usepackage[all]{hypcap}

\usepackage{xcolor}

\newcommand{\nd}{{\phantom\dag}}
\newcommand{\ee}{{\rm e}}

\newcommand{\eps}{\epsilon}
\newcommand{\veps}{\varepsilon}

\newcommand{\vk}{\mathbf{k}}
\renewcommand{\vr}{\mathbf{r}}

\newcommand{\hV}{\hat{V}}
\newcommand{\hW}{\hat{W}}

\hypersetup{
	pdftitle={},
	pdfauthor={},
	colorlinks=true,
	linkcolor=black,
	citecolor=black,
	filecolor=black,
	urlcolor=black,
	bookmarksopen,
	bookmarksopenlevel=1,
}
\begin{document}


\title{Ubiquity of Superconducting Domes in the Bardeen-Cooper-Schrieffer Theory with Finite-Range Potentials}

\author{Edwin Langmann}
\affiliation{Department of Physics, KTH Royal Institute of Technology, SE-106 91 Stockholm, Sweden}
\author{Christopher Triola}
\affiliation{Department of Physics and Astronomy, Uppsala University, Box 516, S-751 20 Uppsala, Sweden}
\author{Alexander V. Balatsky}
\affiliation{Department of Physics, University of Connecticut, Storrs, CT 06269-3046, USA}
\affiliation{Nordic Institute for Theoretical Physics (NORDITA), Stockholm, Sweden}
\affiliation{Center for Quantum Materials (CQM), KTH and Nordita, Stockholm, Sweden}

\date{\today }

\begin{abstract}
Based on recent progress in mathematical physics, we present a reliable method to analytically solve the linearized BCS gap equation for a large class of finite-range interaction potentials leading to s-wave superconductivity. With this analysis, we demonstrate that the monotonic growth of the superconducting critical temperature $T_c$ with the carrier density, $n$, predicted by standard BCS theory, is an artifact of the simplifying assumption that the interaction is quasi-local. In contrast, we show that {\em any} well-defined non-local potential leads to a ``superconducting dome'', i.e.  a non-monotonic $T_c(n)$ exhibiting a maximum value at finite doping and going to zero for large $n$. This proves that, contrary to conventional wisdom, the presence of a superconducting dome is not necessarily an indication of competing orders, nor of exotic superconductivity. 
\end{abstract}


\maketitle


\paragraph*{Introduction.}
It is well-known that the BCS theory of superconductivity \cite{BCS} predicts a critical temperature that increases monotonically with the density of quasiparticles. However, since the discovery of high-temperature superconductors, a growing number of superconducting systems have been revealed to possess critical temperatures, $T_c$,  that have a non-monotonic dependence on either the carrier density or pressure, including: SrTiO$_3$ \cite{koonce1967superconducting,collignon2018metallicity}, the cuprates \cite{dagotto1994correlated,lee2006doping,KKNUZ2015,FKT2015}, the pnictides \cite{FeHTSC}, and heavy fermion superconductors \cite{mathur1998magnetically}. This non-monotonic critical temperature presents itself as a dome of superconductivity in the phase diagram of these systems, and in many cases these domes appear to occur in the neighbourhood of a quantum critical point (QCP) \cite{gegenwart2008quantum,edge2015quantum}. This concomitance is so prevalent that it has resulted in the often quoted rule that beneath every dome there is a QCP of some critical order. In some systems this is likely the case since the presence of a QCP can induce soft bosonic excitations which can act as a ``glue'' leading to the formation of a superconducting state. However, superconducting domes have also been observed in doped band insulators \cite{Ye2011} and magic-angle graphene superlattices \cite{cao2018unconventional} with no sign of competing orders. In these cases a more conventional explanation may be necessary. 

One reason for the incredible success of so many predictions of BCS theory can be attributed to {\em universality}, that is, certain predictions of the theory are independent of model details and thus accurately predicted by simplified models \cite{Leggett}. Famous examples of such universal features of BCS theory include \cite{Tinkham}: the ratio of the superconducting gap at zero temperature to the critical temperature: $\Delta(0)/ T_c \approx 1.76$; and the temperature dependence of the gap for temperatures close to $T_c$: $\Delta(T)/T_c\approx 3.07\sqrt{1-T/T_c}$ (we set $k_B=\hbar=1$ throughout this work). This being said, $T_c$ is non-universal, and accurate predictions of $T_c$ are notoriously difficult; see e.g.\ \cite{AllenMitrovic} for a classic reference and \cite{EKS2018} for a recent discussion. Therefore, it is not clear, a priori, what universal statements can be made about the dependence of $T_c$ on doping or other control parameters. However, recent developments in mathematical physics \cite{HHSS,HS} have significantly improved the mathematical toolkit we can use to extract reliable analytic results for critical temperatures from BCS-like theories.

In this work we take advantage of these recent mathematical insights \cite{HS} to address the general question: when do superconducting domes arise in isotropic BCS models? Surprisingly, we find that superconducting domes arise ubiquitously whenever the electron-electron interaction responsible for the superconductivity has non-trivial spatial dependence and satisfies certain convergence criteria. In this way, we show that the monotonic $T_c$ predicted by BCS theory is actually an artifact of the trivial spatial dependence of the interaction. Furthermore, we present analytic solutions of the linearized gap equation, applicable to a broad class of long-range BCS models, and show that the explicit $T_c$-equation thus obtained is numerically accurate. 

The ubiquity of these domes can be understood to arise from an interplay between the length scale determining the range of interaction, $\ell$, and the average interparticle separation, $\sim k_F^{-1}$. At low densities, when $k_F^{-1}>>\ell$, the interaction becomes effectively local, and the pairing is  well described by standard BCS theory; in this regime, $T_c$ grows as the density increases. Whereas, at high densities, when $\ell>>k_F^{-1}$, the pairing between electrons at the Fermi surface becomes weaker with increasing $k_F$ due to the decay of the interaction in Fourier space,  which suppresses $T_c$ towards zero.  Therefore, in the crossover regime, where $\ell\sim k_F^{-1}$, a superconducting dome arises. This simple explanation of the physics of superconducting domes does not rely on quantum criticality or any other exotic physics. The only necessary ingredient is a mathematically well-behaved electron-electron interaction with nonzero spatial range. 

From a mathematical point of view, our contribution is to extend recent results for $T_c$ \cite{HS} from 0-th order to arbitrary order in a small parameter expansion. This extension is of great importance for the assessment of the numerical accuracy of results obtained using these methods. Additionally, since we use simpler mathematical arguments, we hope that the present paper can act as a bridge between the mathematical physics community working on BCS theory and the broader community of physicists working on superconductivity.

\paragraph*{Generalized BCS model.}
To study the superconducting critical temperature, we employ the standard quantum many-body Hamiltonian  
\begin{multline}
\label{eq:ham}
H = \int \sum_{\sigma=\uparrow,\downarrow}\psi^\dag_{\sigma,\textbf{r}}\left(-\frac{\nabla^2}{2m^*}-\mu\right)\psi_{\sigma,\textbf{r}} d^3r
\\
+ \frac12\iint \sum_{\sigma,\sigma'=\uparrow,\downarrow} \psi^\dag_{\sigma,\textbf{r}}\psi^\dag_{\sigma',\textbf{r}'}V(|\textbf{r}-\textbf{r}'|)\psi^\nd_{\sigma',\textbf{r}'}\psi_{\sigma,\textbf{r}} d^3r d^3r'
\end{multline}
where $\psi^\dag_{\sigma,\textbf{r}}$ ($\psi_{\sigma,\textbf{r}}$) creates (annihilates) a fermion with spin $\sigma$ at position $\textbf{r}$, $\mu$ and $m^*$ are the chemical potential and effective mass, respectively, and $V(r)$ is an attractive non-local interaction potential depending on the interparticle distance $r=|\textbf{r}-\textbf{r}'|$.

The standard textbook BCS model corresponds to the special case where the interaction is quasi-local in position space: $V(\vr)=-g\delta^3(\vr)$ with $g>0$ the coupling strength (we write ``quasi-local'' since the strict local interaction leads to diverging integrals which need to be regularized, as discussed below). We generalize this approach by allowing for finite-range potentials $V(|\vr|)$ of the form
\begin{equation}
\label{VW}
V(r) = -g \ell^{-3} W(r/\ell)
\end{equation}
where $W(x)$ is a function of the dimensionless variable $x=r/\ell\geq 0$ normalized so that $4\pi \int_0^\infty W(x)x^2dx=1$, and $\ell>0$ is the length scale associated with the decay of the interaction in position space. Thus, in the limit $\ell\to 0$, one obtains the textbook BCS model, independent of the function $W(x)$. 

In addition to simple normalization, in this paper we assume that the functions $W(x)$ also satisfy the following two technical conditions 
\cite{HS}:
\begin{equation}
\label{conditions}
\begin{aligned}
(i) & \quad \hW(q)\geq 0,\\
(ii) & \quad \int_0^\infty |W(x)|^p x^2 dx <\infty \mbox{ for } 1\leq p\leq \frac32,
\end{aligned}
\end{equation}
where $\hat{W}(q)=(4\pi/q)\int_0^\infty W(x)\sin(xq)xdx$ is the Fourier transform of $W(x)$. To understand the significance of these conditions, we note that, while the model defined in Eq.~\eqref{eq:ham} for a local potential always leads to s-wave superconductivity, this is not the case for non-local potentials; see \cite{FrankLemm} for counter examples. However, it is known that, if the Fourier transform of the pairing potential, $\hV_{\vk,\vk'}$, is non-positive, one always obtains s-wave superconductivity  \cite{FHNS,HSreview}. This is equivalent to Eq.~\eqref{conditions} (i). Eq.~\eqref{conditions} (ii) guarantees that the BCS gap equation in \eqref{BCSgap} is well-defined \cite{FHNS}, i.e., it rules out potentials that are too singular or which do not decay fast enough at large distances. Some familiar examples of functions which satisfy both of these criteria are:  the Gaussian distribution, the Lorentzian distribution, and the Yukawa potential  (see Table~\ref{table:examples}). 

\paragraph*{$T_c$ for finite-range potentials.}
To find the superconducting critical temperature, $T_c$, associated with the model in Eq.~\eqref{eq:ham}, we will solve the linearized BCS gap equation \cite{Leggett,FHNS}:
\begin{equation}
\label{BCSgap}
\Delta(\eps,T) = -\int\hV(\epsilon,\epsilon')N(\epsilon')
\frac{\tanh{\tfrac{\eps'}{2 T}}}{2\eps'}\Delta(\eps',T)d\eps'
\end{equation}
where $\Delta(\eps)$ is the gap function, depending on energy $\eps=\eps_{\vk}=\tfrac{k^2}{2m^*}-\mu$ and temperature $T$, $N(\eps)=\frac{2m^*}{(2\pi)^2}\theta(1+\eps/\mu)k(\eps)$ is the electronic density of states, and $\hV(\eps,\eps')$ is the average of $\hV_{\vk,\vk'}$ over the energy surfaces $\eps=\eps_{\vk}$ and $\eps'=\eps_{\vk'}$. With these definitions, it is straightforward to show that $\hV(\eps,\eps')$ is given by
\begin{multline}
\label{VN}
\hV(\eps,\eps') =  \theta(1+\eps/\mu)\theta(1+\eps'/\mu)  \\ \times
\frac{ f_W\left(\ell^2[k(\eps)+k(\eps')]^2\right) -  f_W\left(\ell^2[k(\eps)-k(\eps')]^2\right)}{4\ell^2 k(\eps)k(\eps')}
\end{multline}
where $\theta(x)$ is the Heaviside function, $k(\eps)=k_F\sqrt{1+\eps/\mu}$, $k_F=\sqrt{2 m^*\mu}$ is the Fermi momentum, and $f_W(\veps)=\int_0^{\veps}\hW(\sqrt{\veps'})d\veps'$ is a special function determined by the interaction potential. In many cases of interest one can find simple explicit formulas for the function $f_W(\veps)$; see Table~\ref{table:examples} for examples. 

To put this problem in perspective, recall that, for separable potentials of the form $\hV(\eps,\eps')=-g\eta(\eps)\eta(\eps')$, the energy dependence of the gap is trivially determined by the potential: $\Delta(\eps,T)=\Delta(T)\eta(\eps)$. Thus, after insertion into Eq.~\eqref{BCSgap} and cancelling $\Delta(T)$, one obtains an equation involving an integral of known functions and parameters that can be solved for $T$. This was the strategy employed by BCS in their seminal paper \cite{BCS}, using $\eta(\eps)=\theta(\omega_D-|\eps|)$ where $\omega_D>0$ is the Debye energy. As mentioned above, this can be interpreted as Eq.~\eqref{BCSgap} with a local potential and an energy cutoff, $\omega_D$, introduced to regularize a diverging integral. For non-local potentials satisfying Eqs.~\eqref{conditions}, no such ad-hoc regularization is needed: the integral in Eq.~\eqref{BCSgap} is mathematically well-defined \cite{FHNS}. However, the price we must pay is computational difficulty, to solve the gap equation in \eqref{BCSgap} for non-separable potentials one must keep track of the energy dependence of the gap.

Our main result is an explicit formula for $T_c$ in terms of $f_W(\veps)$, obtained by solving Eq.~\eqref{BCSgap} analytically. As we will show in the next section, $T_c$ is given by
\begin{equation}
\label{Tc}
 T_c = \frac{2\ee^{\gamma}}{\pi}\mu \exp\left( -\frac1{\lambda} + a_0 + a_1\lambda + a_2\lambda^2+\cdots \right)
\end{equation}
where $\gamma$ is the Euler-Mascheroni constant, $2\ee^\gamma/\pi\approx 1.13$, the coefficients $a_n$ are given by Eqs.~\eqref{a0} and \eqref{an}, and $\lambda$ is a parameter defined as
\begin{equation}
\label{lambda}
\lambda = -N(0) \hV(0,0)= \frac{2m^*}{(2\pi)^2}k_F g\frac{f_W([2k_F \ell]^2)}{[2 k_F\ell]^2}.
\end{equation}

As explained below, such an explicit formula for $T_c$ can be obtained mainly because the energy scale for superconductivity is exponentially smaller than the chemical potential $\mu$. This is true even in the low-density limit $\mu\to 0$ \cite{HS2}. Indeed, it follows from our result that $T_c/\mu$ goes like $\ee^{-1/\lambda}$ with $\lambda\propto k_F^3 g/\mu$ vanishing like $\sqrt{\mu}$ as $\mu\rightarrow 0$, and if $\lambda$ is sufficiently small, such corrections are negligible. We stress that $\lambda$ can be small even in cases where the coupling strength, $g$, is large. It is an emergent small parameter in the problem whose maximum value occurs at a finite doping such that $k_{F,max}=q_0/2\ell$, where $q_0$ is a numerical value determined only by the form of the interaction potential \cite{SM}.

\paragraph*{Derivation of $T_c$-equation.}
We present our method for solving Eq.~\eqref{BCSgap}. This section can be skipped without loss of continuity if one is only interested in the results. 
 
To solve Eq.~\eqref{BCSgap} we start from the ansatz
\begin{equation}
\label{BCSgap1}
\Delta(\eps) = -\hV(\epsilon,0)N(0)\Delta(0)\log(\Omega_T(\eps)/T)
\end{equation}
which serves as a definition of the function $\Omega_T(\eps)$:
\begin{equation}
\label{OmegaT}
\Omega_T(\eps) =  T \exp\left( \int
\frac{\tanh{\tfrac{\eps'}{2 T}}}{2\eps'}G(\eps,\eps')d\eps'
 \right)
\end{equation}
where  $G(\eps,\eps')=\hV(\epsilon,\epsilon')N(\epsilon')\Delta(\eps')/\hV(\epsilon,0)N(0)\Delta(0)$. In a sense, all we have done is rewrite Eq.~\eqref{BCSgap}; however, using the definition of $\lambda$ in Eq.~\eqref{lambda}, it is clear that $T_c$ is given exactly by $ T_c = \Omega_{T_c}(0)\ee^{-1/\lambda}$. Therefore, the main objective is now to obtain a general method for solving Eq.~\eqref{OmegaT}. 

We now claim that $\Omega_T(\eps)$ has a well-defined limit $T\to 0$, and $\Omega_T(\eps)$ can be replaced by $\Omega_0(\eps)$ up to negligible corrections. To see this, consider the auxiliary quantity $\Omega^{(0)}_T = T\exp\left[\int\frac{\tanh(\eps'/2 T)}{2\eps'}\theta(\mu-|\eps'|) d\eps' \right]$. It is well-known that $\Omega^{(0)}_T \to  \frac{2\ee^{\gamma}}{\pi}\mu$ as $T\to 0$ \cite{BCS,HSreview}, and it is easy to take the limit $T\to 0$ in the ratio $\Omega^\nd_T(\eps)/\Omega^{(0)}_T$. Thus, 
\begin{equation}
\label{OmegaT1}
\Omega_0(\eps) =   \frac{2\ee^{\gamma}}{\pi}\mu  \exp\left( 
\int \frac{  G(\eps,\eps')  -\theta(\mu-|\eps'|)}{2|\eps'|}d\eps'  
 \right), 
\end{equation}
which is well-defined since the integrand remains finite as $\eps'\to 0$. It can be proven that $\log(\Omega_T(\eps)/\Omega_0(\eps))$ vanishes like $(T/\mu)^2$ for small $T/\mu$ \cite{SM}. Since $T/\mu<T_c/\mu$ and $T_c/\mu$ is proportional to $\ee^{-1/\lambda}$, which is negligible for sufficiently small $\lambda$, we can replace $\Omega_T(\eps)$ by $\Omega_0(\eps)$ in the following. 

Inserting Eq.~\eqref{OmegaT1} to Eq.~\eqref{BCSgap1}, for $\eps=0$, we can solve for $T_c$. Ignoring corrections $\propto\ee^{-1/\lambda}$, we find 
\begin{equation} 
\label{Tc0}
 T_c =  \frac{2\ee^{\gamma}}{\pi}\mu  \exp\Biggl\{-\frac1\lambda +
\int \frac{  G(0,\eps')  -\theta(\mu-|\eps'|)}{2|\eps'|}d\eps' 
 \Biggr\}. 
\end{equation} 
To compute $G(0,\eps')$ we use 
\begin{equation} 
\frac{\Delta(\eps)}{\Delta(0)} = \frac{\hV(\epsilon,0)}{\hV(0,0)}\left(1 - \lambda \log\left[\frac{\Omega_{T_c}(\eps)}{\Omega_{T_c}(0)}\right]\right) 
\end{equation} 
which follows from Eq.~\eqref{BCSgap1} and the exact implicit $T_c$-equation above by straightforward computations. Replacing $ \log\left[\Omega_{T_c}(\eps)/\Omega_{T_c}(0)\right]$ by $\log\left[\Omega_{0}(\eps)/\Omega_{0}(0)\right]$, ignoring corrections $\propto\ee^{-1/\lambda}$, we can write this as
\begin{equation} 
\frac{\Delta(\eps)}{\Delta(0)} = \frac{\hV(\epsilon,0)}{\hV(0,0)} + \lambda \int K(\eps,\eps') \frac{\Delta(\eps')}{\Delta(0)} d\eps'
\end{equation} 
with the integral kernel 
\begin{equation} 
\label{K}
K(\eps,\eps') = \frac{\hV(\eps,0)}{\hV(0,0)}\frac1{2|\eps'|}\Biggl[  \frac{\hV(\eps,\eps')}{\hV(\eps,0)} -\frac{\hV(0,\eps')}{\hV(0,0)} 
 \Biggr] \frac{N(\eps')}{N(0)}. 
\end{equation} 
This is an inhomogeneous Fredholm integral equation which can be solved by iteration: $\Delta(\eps)/\Delta(0) = F_0(\eps)+F_1(\eps)\lambda+F_2(\eps)\lambda^2+\cdots$ with 
\begin{equation} 
\label{Fn} 
F_0(\eps)= \frac{\hV(\eps,0)}{\hV(0,0)},\quad F_{n}(\eps)=\int K(\eps,\eps')F_{n-1}(\eps')d\eps'
\end{equation} 
for $n=1,2,\ldots$. Recalling the definition of $G(0,\eps')$ we use Eq.~\eqref{Tc0} to obtain Eq.~\eqref{Tc} with 
\begin{equation}
\label{a0}
a_0 = \int \frac{1}{2|\eps'|}\,  \left( \frac{\hV(0,\eps')N(\eps')}{\hV(0,0) N(0)}F_0(\eps') - \theta(\mu-|\eps'|) \right) d\eps'
\end{equation} 
and 
\begin{equation} 
\label{an}
a_n =  \int \frac{\hV(0,\epsilon')N(\epsilon')}{\hV(0,0)N(0)} \frac{F_n(\eps')}{2|\eps'|}d\eps'
\end{equation}  
for $n=1,2,\ldots$ (note that all $a_{n\geq 0}$ are well-defined since the integrands in \eqref{a0} and \eqref{an} remain finite as $\eps'\to 0$). 

\begin{figure}
 \begin{center}
  \centering
\includegraphics[width=0.45\textwidth]{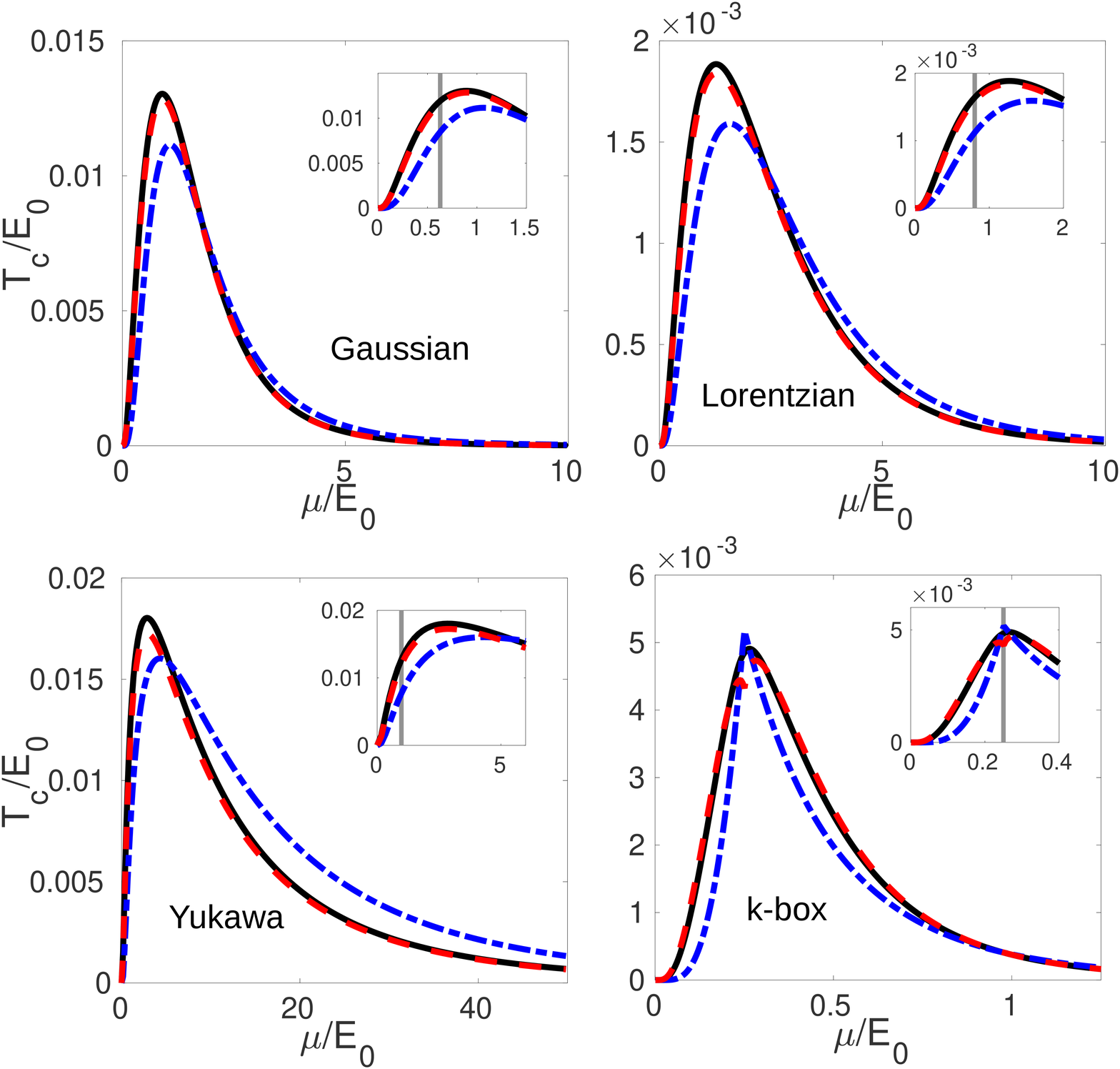}
\caption{Plots of the critical temperature $T_c$ as a function of the chemical potential $\mu$ for the finite-range potential examples given in Table \ref{table:examples}. In each case we present: (blue/dashed-dotted) $T_c^{(-1)}$, computed by truncating Eq.~\eqref{Tc} at the $1/\lambda$ order; (red/dashed) $T_c^{(0)}$, computed by truncating Eq.~\eqref{Tc} at the $a_0$ order; and (black/solid) in which the 1-st order correction $a_1\lambda$ is included. In each case, the latter two curves agree remarkably well, and the first is a reasonable approximation capturing the qualitative behavior. All energies are reported in units of $E_0=1/2m^*\ell^2$ where $\ell$ is the interaction range, and the coupling constant is set so that $m^* g/(2\pi)^2\ell=0.5$. The insets show the small $\mu$ (low concentration) behavior of the respective $T_c$ with the vertical line indicating the $\mu$-value at which $\lambda$ has its maximum.
}
\end{center}
\label{fig:examples}
\end{figure}

\paragraph*{Superconducting domes} 
To gain some insight into the universal properties of the formula for $T_c$, Eq.~\eqref{Tc}, in Fig.~\ref{fig:examples} we plot $ T_c$ as a function of the chemical potential, $\mu$, \cite{nt} for the four examples appearing in Table~\ref{table:examples}, using the same coupling constant $g$ for each case. While the spatial dependence of the interaction, $W(x)$, obviously has a large effect on the finer structure of each phase diagram, clearly, all four examples exhibit superconducting domes. These domes appear, despite the fact that the examples possess wildly different spatial dependence, for example: the Lorentzian distribution decays much more slowly in space, and the ``$k$-box'' potential is actually oscillatory in space. 

\begin{table}[htb]
\begin{tabular}{ l | l | l | l }
\hline\hline
& $W(x)$ & $\hat{W}(q)$ & $f_W(\veps)$   \\ [0.6ex]
\hline
Gaussian& $\frac1{(2\pi)^{3/2}}\ee^{-x^2/2}$ & $\ee^{-q^2/2}$ & $2\left(1-\ee^{-\veps/2} \right)$ \\ [0.6ex]
\hline
Lorentzian & $\frac1{\pi^2(1+x^2)^2}$  & $\ee^{-|q|}$ & $2\left[1-\ee^{-\sqrt{\veps}}\left(1+\sqrt{\veps}\right)\right]$ \\ [0.6ex]
\hline
Yukawa & $\frac{1}{4\pi x}\ee^{-x}$ & $\frac1{1+q^2}$ & $\ln(1+\veps)$ \\ [0.6ex]
\hline
$k$-box \cite{nt1} & $\frac{\sin(x)-x\cos(x)}{2\pi^2x^3}$ & $\theta(1-|q|)$ & $\veps\theta(1-\veps)+\theta(\veps-1)$ \\ [0.6ex]
\hline
\end{tabular}
\caption{Examples of functions $W(x)$ determining finite-range potentials as in Eq.~\eqref{VW}, together with their Fourier transforms $\hat{W}(q)$ and associated functions $f_W(\veps)$.}
\label{table:examples}
\end{table}

The emergence of these superconducting domes is a direct consequence of the doping-dependence of $\lambda$ which is the product of two factors, the density of states at the Fermi level, $N(0)$, and the interaction strength between quasiparticles at the Fermi level, $\hat{V}(0,0)$. Each of these factors have different doping-dependences, while $N(0)$ increases monotonically with doping, $\hat{V}(0,0)$ gets weaker at large doping, due to decay of the Fourier coefficients of the interaction potential at large momenta. This can be understood more rigorously by considering the second equality in Eq.~\eqref{lambda} which implies $\lambda\propto m^*g \ell^{-1} f_W(\veps)/\sqrt{\veps}$ with $\veps=(2k_F\ell)^2=4\mu/E_0$. One can show that $f_W(\veps)/\sqrt{\veps}\to 0$ in both limits $\veps\to 0$ and $\veps\to\infty$ provided the condition in Eq.~\eqref{conditions} (ii) holds true. Since the behavior of $T_c$ is dominated by the factor $\ee^{-1/\lambda}$ for small $\lambda$, this implies that $T_c$ vanishes both in the low- and high-density limits, as described in the introduction. Therefore, we conclude that {\em superconducting domes are ubiquitous in BCS theory with finite-range potentials such that the BCS gap equation is well-defined.}

It is interesting to note that the vanishing of $T_c$ in the large-density limit is related to the short-distance behavior of the potential $V(r)$. In order for the ratio $f_W(\veps)/\sqrt{\veps}$ to not approach zero as $\veps \to \infty$, $V(r)$ must have a $1/r^\alpha$-singularity at $r\to 0$ with $\alpha\geq 2$. However, such singular potentials violate Eq.~\eqref{conditions} (ii), and, thus, the BCS equation in \eqref{BCSgap} is not mathematically well-defined. This means that, for well-defined potentials with finite spatial range, the pairing between electrons at the Fermi surface becomes weaker at high-doping due to the decay of the interaction in Fourier space.

While superconducting domes are ubiquitous, the precise spatial dependence of the potential can be responsible for some significant features in the finer structure of the phase diagram. For example, in Fig.~\ref{fig:examples}, we see that the different potentials give rise to critical temperatures of very different energy scales, differing by orders of magnitude for the four examples. This demonstrates another key point of our results: it is not just the coupling strength which determines the magnitude of $T_c$, the precise form of the interaction potential can make a huge difference. 
Additionally, we note that, in the case of the $k$-box potential, if we approximate the $T_c$ formula by truncating the series in Eq.~\eqref{Tc} at order $-1$ or $0$ the phase diagram in Fig.~\ref{fig:examples} exhibits a cusp, resulting from the oscillatory spatial dependence of this potential. However, this cusp disappears at order 1, illustrating the point that, for certain potentials, the higher order corrections can be significant even for weak coupling.

\paragraph*{Conclusions}
BCS theory and its generalization due to Eliashberg have provided a remarkably successful theoretical framework to describe many superconducting materials. Still, some properties of superconductors have remained difficult to explain from first principles, or even from a knowledge of the normal state. One such property is the superconducting critical temperature $T_c$. In this paper we have called attention to one microscopic detail that has limited the accuracy of $T_c$-predictions: the spatial dependence of the pairing interaction. We presented a reliable method to take this into account in BCS theory, and we demonstrated its importance, both quantitatively and qualitatively. 

Importantly, we showed that a non-trivial spatial dependence of the pairing interaction in BCS theory leads to superconducting domes. While the exact scale of $T_c$ depends sensitively on interaction details, we found a wide class of ``reasonable'' potentials which induce domes with $T_c(n)$-dependences that look remarkably similar to one another upon scaling. 

The superconducting domes we find are controlled by the ratio of interparticle distance to the effective range of the potential, and they do not rely on any competing order or quantum critical fluctuations of the competing phases in the vicinity of superconducting state. These findings give greater confidence to the applicability of standard BCS results and establish the interesting possibility that superconducting domes can occur in simple BCS superconductors with no competing orders. 

\begin{acknowledgments}
\paragraph*{Acknowledgments.}
We thank Kamran Behnia, Annica M. Black-Schaffer, G\"oran Grimvall, Christian Hainzl, Yaron Kedem, Tomas L\"{o}thman, Andreas Rydh, and Robert Seiringer for helpful discussions. We are grateful to Kamran Behnia and Andreas Rydh for useful comments on the manuscript. We also would like to thank the referees for valuable suggestions which helped us to improve this paper. 
E.L.\ acknowledges support from the Swedish Research Council (VR Grant No.\ 2016-05167). The work of A. V. B. is supported by the Knut and Alice Wallenberg Foundation, the Swedish Research Council (VR Grant No. 2017-03997), and the Villum Fonden via the Centre of Excellence for Dirac Materials (Grant No. 11744). The work of C.T. was supported by the Swedish Research Council (VR Grant No. 621-2014-3721).
\end{acknowledgments}

\bibliographystyle{apsrev}
\bibliography{sc_domes}

\begin{thebibliography}{26}
\expandafter\ifx\csname natexlab\endcsname\relax\def\natexlab#1{#1}\fi
\expandafter\ifx\csname bibnamefont\endcsname\relax
  \def\bibnamefont#1{#1}\fi
\expandafter\ifx\csname bibfnamefont\endcsname\relax
  \def\bibfnamefont#1{#1}\fi
\expandafter\ifx\csname citenamefont\endcsname\relax
  \def\citenamefont#1{#1}\fi
\expandafter\ifx\csname url\endcsname\relax
  \def\url#1{\texttt{#1}}\fi
\expandafter\ifx\csname urlprefix\endcsname\relax\def\urlprefix{URL }\fi
\providecommand{\bibinfo}[2]{#2}
\providecommand{\eprint}[2][]{\url{#2}}

\bibitem[{\citenamefont{Bardeen et~al.}(1957)\citenamefont{Bardeen, Cooper, and
  Schrieffer}}]{BCS}
\bibinfo{author}{\bibfnamefont{J.}~\bibnamefont{Bardeen}},
  \bibinfo{author}{\bibfnamefont{L.~N.} \bibnamefont{Cooper}},
  \bibnamefont{and} \bibinfo{author}{\bibfnamefont{J.~R.}
  \bibnamefont{Schrieffer}}, \bibinfo{journal}{Phys. Rev.}
  \textbf{\bibinfo{volume}{108}}, \bibinfo{pages}{1175} (\bibinfo{year}{1957}).

\bibitem[{\citenamefont{Koonce et~al.}(1967)\citenamefont{Koonce, Cohen,
  Schooley, Hosler, and Pfeiffer}}]{koonce1967superconducting}
\bibinfo{author}{\bibfnamefont{C.}~\bibnamefont{Koonce}},
  \bibinfo{author}{\bibfnamefont{M.~L.} \bibnamefont{Cohen}},
  \bibinfo{author}{\bibfnamefont{J.}~\bibnamefont{Schooley}},
  \bibinfo{author}{\bibfnamefont{W.}~\bibnamefont{Hosler}}, \bibnamefont{and}
  \bibinfo{author}{\bibfnamefont{E.}~\bibnamefont{Pfeiffer}},
  \bibinfo{journal}{Phys. Rev.} \textbf{\bibinfo{volume}{163}},
  \bibinfo{pages}{380} (\bibinfo{year}{1967}).

\bibitem[{\citenamefont{Collignon et~al.}(2019)\citenamefont{Collignon, Lin,
  Rischau, Fauqu{\'e}, and Behnia}}]{collignon2018metallicity}
\bibinfo{author}{\bibfnamefont{C.}~\bibnamefont{Collignon}},
  \bibinfo{author}{\bibfnamefont{X.}~\bibnamefont{Lin}},
  \bibinfo{author}{\bibfnamefont{C.~W.} \bibnamefont{Rischau}},
  \bibinfo{author}{\bibfnamefont{B.}~\bibnamefont{Fauqu{\'e}}},
  \bibnamefont{and} \bibinfo{author}{\bibfnamefont{K.}~\bibnamefont{Behnia}},
  \bibinfo{journal}{Annual Review of Condensed Matter Physics}
  (\bibinfo{year}{2019}).

\bibitem[{\citenamefont{Dagotto}(1994)}]{dagotto1994correlated}
\bibinfo{author}{\bibfnamefont{E.}~\bibnamefont{Dagotto}},
  \bibinfo{journal}{Rev. Mod. Phys.} \textbf{\bibinfo{volume}{66}},
  \bibinfo{pages}{763} (\bibinfo{year}{1994}).

\bibitem[{\citenamefont{Lee et~al.}(2006)\citenamefont{Lee, Nagaosa, and
  Wen}}]{lee2006doping}
\bibinfo{author}{\bibfnamefont{P.~A.} \bibnamefont{Lee}},
  \bibinfo{author}{\bibfnamefont{N.}~\bibnamefont{Nagaosa}}, \bibnamefont{and}
  \bibinfo{author}{\bibfnamefont{X.-G.} \bibnamefont{Wen}},
  \bibinfo{journal}{Rev. Mod. Phys.} \textbf{\bibinfo{volume}{78}},
  \bibinfo{pages}{17} (\bibinfo{year}{2006}).

\bibitem[{\citenamefont{Keimer et~al.}(2015)\citenamefont{Keimer, Kivelson,
  Norman, Uchida, and Zaanen}}]{KKNUZ2015}
\bibinfo{author}{\bibfnamefont{B.}~\bibnamefont{Keimer}},
  \bibinfo{author}{\bibfnamefont{S.~A.} \bibnamefont{Kivelson}},
  \bibinfo{author}{\bibfnamefont{M.~R.} \bibnamefont{Norman}},
  \bibinfo{author}{\bibfnamefont{S.}~\bibnamefont{Uchida}}, \bibnamefont{and}
  \bibinfo{author}{\bibfnamefont{J.}~\bibnamefont{Zaanen}},
  \bibinfo{journal}{Nature} \textbf{\bibinfo{volume}{518}},
  \bibinfo{pages}{179} (\bibinfo{year}{2015}).

\bibitem[{\citenamefont{Fradkin et~al.}(2015)\citenamefont{Fradkin, Kivelson,
  and Tranquada}}]{FKT2015}
\bibinfo{author}{\bibfnamefont{E.}~\bibnamefont{Fradkin}},
  \bibinfo{author}{\bibfnamefont{S.~A.} \bibnamefont{Kivelson}},
  \bibnamefont{and} \bibinfo{author}{\bibfnamefont{J.~M.}
  \bibnamefont{Tranquada}}, \bibinfo{journal}{Rev. Mod. Phys.}
  \textbf{\bibinfo{volume}{87}}, \bibinfo{pages}{457} (\bibinfo{year}{2015}).

\bibitem[{\citenamefont{Shibauchi et~al.}(2014)\citenamefont{Shibauchi,
  Carrington, and Matsuda}}]{FeHTSC}
\bibinfo{author}{\bibfnamefont{T.}~\bibnamefont{Shibauchi}},
  \bibinfo{author}{\bibfnamefont{A.}~\bibnamefont{Carrington}},
  \bibnamefont{and} \bibinfo{author}{\bibfnamefont{Y.}~\bibnamefont{Matsuda}},
  \bibinfo{journal}{Annu. Rev. Condens. Matter Phys.}
  \textbf{\bibinfo{volume}{5}}, \bibinfo{pages}{113} (\bibinfo{year}{2014}).

\bibitem[{\citenamefont{Mathur et~al.}(1998)\citenamefont{Mathur, Grosche,
  Julian, Walker, Freye, Haselwimmer, and Lonzarich}}]{mathur1998magnetically}
\bibinfo{author}{\bibfnamefont{N.}~\bibnamefont{Mathur}},
  \bibinfo{author}{\bibfnamefont{F.}~\bibnamefont{Grosche}},
  \bibinfo{author}{\bibfnamefont{S.}~\bibnamefont{Julian}},
  \bibinfo{author}{\bibfnamefont{I.}~\bibnamefont{Walker}},
  \bibinfo{author}{\bibfnamefont{D.}~\bibnamefont{Freye}},
  \bibinfo{author}{\bibfnamefont{R.}~\bibnamefont{Haselwimmer}},
  \bibnamefont{and}
  \bibinfo{author}{\bibfnamefont{G.}~\bibnamefont{Lonzarich}},
  \bibinfo{journal}{Nature} \textbf{\bibinfo{volume}{394}}, \bibinfo{pages}{39}
  (\bibinfo{year}{1998}).

\bibitem[{\citenamefont{Gegenwart et~al.}(2008)\citenamefont{Gegenwart, Si, and
  Steglich}}]{gegenwart2008quantum}
\bibinfo{author}{\bibfnamefont{P.}~\bibnamefont{Gegenwart}},
  \bibinfo{author}{\bibfnamefont{Q.}~\bibnamefont{Si}}, \bibnamefont{and}
  \bibinfo{author}{\bibfnamefont{F.}~\bibnamefont{Steglich}},
  \bibinfo{journal}{Nat. Phy.} \textbf{\bibinfo{volume}{4}},
  \bibinfo{pages}{186} (\bibinfo{year}{2008}).

\bibitem[{\citenamefont{Edge et~al.}(2015)\citenamefont{Edge, Kedem, Aschauer,
  Spaldin, and Balatsky}}]{edge2015quantum}
\bibinfo{author}{\bibfnamefont{J.~M.} \bibnamefont{Edge}},
  \bibinfo{author}{\bibfnamefont{Y.}~\bibnamefont{Kedem}},
  \bibinfo{author}{\bibfnamefont{U.}~\bibnamefont{Aschauer}},
  \bibinfo{author}{\bibfnamefont{N.~A.} \bibnamefont{Spaldin}},
  \bibnamefont{and} \bibinfo{author}{\bibfnamefont{A.~V.}
  \bibnamefont{Balatsky}}, \bibinfo{journal}{Phys. Rev. Lett.}
  \textbf{\bibinfo{volume}{115}}, \bibinfo{pages}{247002}
  (\bibinfo{year}{2015}).

\bibitem[{\citenamefont{Ye et~al.}(2012)\citenamefont{Ye, Zhang, Akashi,
  Bahramy, Arita, and Iwasa}}]{Ye2011}
\bibinfo{author}{\bibfnamefont{J.~T.} \bibnamefont{Ye}},
  \bibinfo{author}{\bibfnamefont{Y.~J.} \bibnamefont{Zhang}},
  \bibinfo{author}{\bibfnamefont{R.}~\bibnamefont{Akashi}},
  \bibinfo{author}{\bibfnamefont{M.~S.} \bibnamefont{Bahramy}},
  \bibinfo{author}{\bibfnamefont{R.}~\bibnamefont{Arita}}, \bibnamefont{and}
  \bibinfo{author}{\bibfnamefont{Y.}~\bibnamefont{Iwasa}},
  \bibinfo{journal}{Science} \textbf{\bibinfo{volume}{338}},
  \bibinfo{pages}{1193} (\bibinfo{year}{2012}).

\bibitem[{\citenamefont{Cao et~al.}(2018)\citenamefont{Cao, Fatemi, Fang,
  Watanabe, Taniguchi, Kaxiras, and Jarillo-Herrero}}]{cao2018unconventional}
\bibinfo{author}{\bibfnamefont{Y.}~\bibnamefont{Cao}},
  \bibinfo{author}{\bibfnamefont{V.}~\bibnamefont{Fatemi}},
  \bibinfo{author}{\bibfnamefont{S.}~\bibnamefont{Fang}},
  \bibinfo{author}{\bibfnamefont{K.}~\bibnamefont{Watanabe}},
  \bibinfo{author}{\bibfnamefont{T.}~\bibnamefont{Taniguchi}},
  \bibinfo{author}{\bibfnamefont{E.}~\bibnamefont{Kaxiras}}, \bibnamefont{and}
  \bibinfo{author}{\bibfnamefont{P.}~\bibnamefont{Jarillo-Herrero}},
  \bibinfo{journal}{Nature} \textbf{\bibinfo{volume}{556}}, \bibinfo{pages}{43}
  (\bibinfo{year}{2018}).

\bibitem[{\citenamefont{Leggett}(2006)}]{Leggett}
\bibinfo{author}{\bibfnamefont{A.~J.} \bibnamefont{Leggett}},
  \emph{\bibinfo{title}{Quantum liquids: Bose condensation and {C}ooper pairing
  in condensed-matter systems}} (\bibinfo{publisher}{Oxford university press},
  \bibinfo{year}{2006}).

\bibitem[{\citenamefont{Tinkham}(1996)}]{Tinkham}
\bibinfo{author}{\bibfnamefont{M.}~\bibnamefont{Tinkham}},
  \emph{\bibinfo{title}{Introduction to superconductivity}}
  (\bibinfo{publisher}{McGraw-Hill, Inc.}, \bibinfo{year}{1996}).

\bibitem[{\citenamefont{Allen and Mitrovi{\'c}}(1983)}]{AllenMitrovic}
\bibinfo{author}{\bibfnamefont{P.~B.} \bibnamefont{Allen}} \bibnamefont{and}
  \bibinfo{author}{\bibfnamefont{B.}~\bibnamefont{Mitrovi{\'c}}},
  \bibinfo{journal}{Sol. Stat. Phys.} \textbf{\bibinfo{volume}{37}},
  \bibinfo{pages}{1} (\bibinfo{year}{1983}).

\bibitem[{\citenamefont{Esterlis et~al.}(2018)\citenamefont{Esterlis, Kivelson,
  and Scalapino}}]{EKS2018}
\bibinfo{author}{\bibfnamefont{I.}~\bibnamefont{Esterlis}},
  \bibinfo{author}{\bibfnamefont{S.}~\bibnamefont{Kivelson}}, \bibnamefont{and}
  \bibinfo{author}{\bibfnamefont{D.}~\bibnamefont{Scalapino}},
  \bibinfo{journal}{npj Quantum Materials} \textbf{\bibinfo{volume}{3}},
  \bibinfo{pages}{59} (\bibinfo{year}{2018}).

\bibitem[{\citenamefont{Hainzl et~al.}(2008)\citenamefont{Hainzl, Hamza,
  Seiringer, and Solovej}}]{HHSS}
\bibinfo{author}{\bibfnamefont{C.}~\bibnamefont{Hainzl}},
  \bibinfo{author}{\bibfnamefont{E.}~\bibnamefont{Hamza}},
  \bibinfo{author}{\bibfnamefont{R.}~\bibnamefont{Seiringer}},
  \bibnamefont{and} \bibinfo{author}{\bibfnamefont{J.~P.}
  \bibnamefont{Solovej}}, \bibinfo{journal}{Comm. Math. Phys.}
  \textbf{\bibinfo{volume}{281}}, \bibinfo{pages}{349} (\bibinfo{year}{2008}).

\bibitem[{\citenamefont{Hainzl and Seiringer}(2008{\natexlab{a}})}]{HS}
\bibinfo{author}{\bibfnamefont{C.}~\bibnamefont{Hainzl}} \bibnamefont{and}
  \bibinfo{author}{\bibfnamefont{R.}~\bibnamefont{Seiringer}},
  \bibinfo{journal}{Phys. Rev. B} \textbf{\bibinfo{volume}{77}},
  \bibinfo{pages}{184517} (\bibinfo{year}{2008}{\natexlab{a}}).

\bibitem[{\citenamefont{Frank and Lemm}(2016)}]{FrankLemm}
\bibinfo{author}{\bibfnamefont{R.~L.} \bibnamefont{Frank}} \bibnamefont{and}
  \bibinfo{author}{\bibfnamefont{M.}~\bibnamefont{Lemm}},
  \bibinfo{journal}{Ann. H. Poincar{\'e}} \textbf{\bibinfo{volume}{17}},
  \bibinfo{pages}{2285} (\bibinfo{year}{2016}).

\bibitem[{\citenamefont{Frank et~al.}(2007)\citenamefont{Frank, Hainzl, Naboko,
  and Seiringer}}]{FHNS}
\bibinfo{author}{\bibfnamefont{R.~L.} \bibnamefont{Frank}},
  \bibinfo{author}{\bibfnamefont{C.}~\bibnamefont{Hainzl}},
  \bibinfo{author}{\bibfnamefont{S.}~\bibnamefont{Naboko}}, \bibnamefont{and}
  \bibinfo{author}{\bibfnamefont{R.}~\bibnamefont{Seiringer}},
  \bibinfo{journal}{J. Geom. Anal.} \textbf{\bibinfo{volume}{17}},
  \bibinfo{pages}{559} (\bibinfo{year}{2007}).

\bibitem[{\citenamefont{Hainzl and Seiringer}(2016)}]{HSreview}
\bibinfo{author}{\bibfnamefont{C.}~\bibnamefont{Hainzl}} \bibnamefont{and}
  \bibinfo{author}{\bibfnamefont{R.}~\bibnamefont{Seiringer}},
  \bibinfo{journal}{J. Math. Phys.} \textbf{\bibinfo{volume}{57}},
  \bibinfo{pages}{021101} (\bibinfo{year}{2016}).

\bibitem[{\citenamefont{Hainzl and Seiringer}(2008{\natexlab{b}})}]{HS2}
\bibinfo{author}{\bibfnamefont{C.}~\bibnamefont{Hainzl}} \bibnamefont{and}
  \bibinfo{author}{\bibfnamefont{R.}~\bibnamefont{Seiringer}},
  \bibinfo{journal}{Lett. Math. Phys.} \textbf{\bibinfo{volume}{84}},
  \bibinfo{pages}{99} (\bibinfo{year}{2008}{\natexlab{b}}).

\bibitem[{SM()}]{SM}
\bibinfo{howpublished}{Supplemental Material (appended below) containing some
  mathematical details used in our derivation of the $T_c$ equation, together
  with numerical results exploring the accuracy of various approximations to
  the full $T_c$ equation.}

\bibitem[{nt()}]{nt}
\bibinfo{howpublished}{We find it convenient to plot $T_c$ as a function of the
  chemical potential $\mu$. It is easy to convert $\mu$ into $n$ if $T_c$ is
  much smaller than all other energy scales: one then can identify $\mu$ with
  $k_F^2/2m^*$ and $n$ with $4\pi k_F^3/3$ (volume of Fermi sphere), and thus
  $n/n_0 = (\mu/E_0)^{3/2}$. For larger $T_c$ values there is a more
  complicated relation between $\mu$ and $n$.}

\bibitem[{nt1()}]{nt1}
\bibinfo{howpublished}{This example decays only like $1/r^2$ as $r\to\infty$
  and thus violates Eq.~\eqref{conditions} (ii). However, we have checked that
  our result is still well-defined and numerically accurate. This is not the
  case for potentials which are more singular for $r\to 0$ than allowed by
  Eq.~\eqref{conditions} (ii).}

\end{thebibliography}

\pagebreak
\onecolumngrid

\begin{center}
\textbf{%
Supplemental Material for:\\ 
Ubiquity of Superconducting Domes in the Bardeen-Cooper-Schrieffer Theory with Finite-Range Potentials}%
\end{center}

\bigskip

This supplement elaborates on some details which were omitted from the main text. In Section~1 we sketch some of the steps in our derivation of the $T_c$-equation presented in the main text. In Section~2 we present additional numerical results studying the validity of some approximations to the full $T_c$-equation for various coupling strengths and doping levels.

\subsection*{1. Temperature dependence of $\Omega_T(\eps)$}
In this section we show by direct computation that the function $\Omega_T(\eps)$, Eq.(9) of the main text, satisfies 
\begin{equation} 
\label{result}
\log\left(\frac{\Omega_T(\eps)}{\Omega_0(\eps)} \right) =\left. -\frac{\pi^2}{24} T^2 \frac{\partial^2}{\partial(\eps')^2}G(\eps,\eps')\right|_{\eps'=0} + \ldots
\end{equation} 
with the dots indicating higher-order terms in $T$. Since the variable $\eps'$ in $G(\eps,\eps')$ scales with $\mu$, this computation proves that $\log\left(\Omega_T(\eps)/\Omega_0(\eps)\right)$ vanishes like $(T/\mu)^2$ for small $T/\mu$, as claimed in the main text. 

To begin, we write Eq.~(9) as 
\begin{equation} 
\Omega_T= T\exp(I(T)),\quad I(T) =  \int G(\eps')\frac{\tanh\frac{\eps'}{2T}}{2\eps'}d\eps'
\label{eq:definition} 
\end{equation} 
where $G(\eps')$ is shorthand for $G(\eps,\eps')$, and we have suppresed the $\eps$-dependence of $\Omega_T$ and $I(T)$, for brevity. We then compute 
\begin{equation} 
\frac{dI(T)}{dT} = -\frac1{4T^2}\int G(\eps') \frac1{\cosh^2\frac{\eps'}{2T}}d\eps' = -\frac1{2T}\int G(2Ty) \frac1{\cosh^2(y)}dy
\end{equation} 
changing the integration variable to $y=\eps'/2T$ in the last step. Inserting the Taylor series $G(\eps')=G(0)+G'(0)\eps'+\frac12 G''(0)(\eps')^2+\ldots $ and recalling that $G(0)=1$, we obtain 
\begin{equation} 
\label{Fp} 
\frac{dI(T)}{dT} = -\frac1{2T}\int \left(1 + G'(0)2Ty + G''(0)2T^2y^2+\ldots  \right)\frac1{\cosh^2(y)}dy = -\frac1T  -\frac{\pi^2}{12} G''(0) T + \cdots , 
\end{equation} 
which implies 
 \begin{equation} 
 \label{F1} 
I(T) = \log(\Omega_0/T) -\frac{\pi^2}{24} G''(0) T^2 + \cdots 
\end{equation} 
with $\Omega_0$ arising as an integration constant. Combining Eq.~\eqref{F1} with the definition of $I(T)$ in Eq.~\eqref{eq:definition} we obtain Eq.~\eqref{result}. 

\subsection*{2. Numerical study of approximations for $T_c$}
Our result for $T_c$ in Eq.(6) is valid to all orders in $\lambda$ but, in practice, it is convenient to truncate the series and approximate $T_c$ as follows, 
\begin{equation}
\label{Tc1}
 T_c^{(-1)} = \frac{2\ee^{\gamma}}{\pi}\mu \exp\left( -\frac1{\lambda} \right),\quad 
 T^{(n)}_c = \frac{2\ee^{\gamma}}{\pi}\mu \exp\left( -\frac1{\lambda} + \sum^n_{m=0}a_m \lambda^m \right), \ n\geq 0.
\end{equation} 
In this section we will evaluate $T_n^{(-1)}$, $T_n^{(0)}$, and $T_n^{(1)}$ numerically, and illustrate how these three approximations depend on the size of the BCS coupling parameter, $\lambda$. Recall that $\lambda$ is proportional to the bare coupling constant, $g$, but has a non-monotonic dependence on the Fermi wavevector, $k_F$, therefore, it will be convenient to fix one of these parameters as we vary the other. 

The value of $k_F$ for which $\lambda$ has its maxiumum can be determined from 
\begin{equation} 
\label{lambda2} 
\lambda = \frac{g}{g_0}\frac{f_W([2k_F\ell]^2)}{2k_F\ell},\quad g_0 = \frac{(2\pi)^2\ell}{m^*}. 
\end{equation} 
Using this it is straightforward to show that the $\lambda(k_F)$-maximum occurs at $k_F=k_{F,max}=q_0/2\ell$ where $q_0$ satisfies 
\begin{equation}
\label{Max}
2q_0^2\frac{f'_W\left(q_0^2\right)}{f_W\left(q_0^2 \right)}=1,
\end{equation}
($f'_W\left(\veps\right) = \partial f_W(\veps)/\partial \veps$).  Importantly, the value of $q_0$ is independent of $g$ and all other parameters of the model: it is a characteristic of the normalized functions describing the spatial dependence of the interaction potential, $W(x)$. 

It is straightforward to solve Eq.~\eqref{Max} for specific examples, making sure that the solution maximizes $\lambda(k_F)$. In Table~\ref{table:examples2} we give these solutions $q_0=2\ell k_{F,max}$  for the four example potentials in Table~I, together with the values of $\tilde\lambda(k_{F})\equiv f_W([2k_F\ell]^2)/2k_F\ell$ for $k_F=k_{F,max}/2$, $k_{F,max}$ and $2k_{F,max}$, which can be used to compute the corresponding 
values of $\lambda(k_F)$ using Eq.~\eqref{lambda2}. For the case of the $k$-box potential, the solution could be found analytically; however, for the other three cases we proceeded numerically.  

\begin{table}[htb]
\begin{tabular}{ l | l | l | l | l | l }
\hline\hline
& $f_W(\varepsilon)$ & $q_0$ & $\tilde{\lambda}(k_{F,max}/2)$ & $\tilde{\lambda}(k_{F,max})$ & $\tilde{\lambda}(2k_{F,max})$   \\ [0.6ex]
\hline
Gaussian& $2\left(1-e^{-\varepsilon/2} \right)$ & $1.5852...$ &  $0.6802...$ & $0.9025...$ & $0.6267...$  \\ [0.6ex]
\hline
Lorentzian & $2\left[1-e^{-\sqrt{\varepsilon}}\left(1+\sqrt{\varepsilon}\right)\right]$  & $1.7932...$  & $0.5047...$ & $0.5969...$ & $0.4868...$ \\ [0.6ex]
\hline
Yukawa & $\ln(1+\varepsilon)$ & $1.9803...$  & $0.6901...$ & $0.8047...$ & $0.7106...$ \\ [0.6ex]
\hline
$k$-box & $\varepsilon\theta(1-\varepsilon)+\theta(\varepsilon-1)$ & $1$ &  $0.5$ & $1$ & $0.5$ \\ [0.6ex]
\hline
\end{tabular}
\caption{(Continuation of Table~I.) 
Numerical values for the solutions, $q_0$, to Eq.~(\ref{Max}) for the four examples of interaction potentials given in Table~I. As explained in the text, $q_0$ determines the Fermi wavevector associated with the maximum value of $\lambda$, $k_{F,max}=q_0/2\ell$. In each case we have also included numerical values for $\tilde{\lambda}(k_{F})=\lambda(k_{F})g_0/g$ for $k_F=k_{F,max}/2$, $k_{F,max}$ and $2k_{F,max}$.  
} 
\label{table:examples2}
\end{table}

\begin{figure}
 \begin{center}
  \centering
\includegraphics[width=0.8\textwidth]{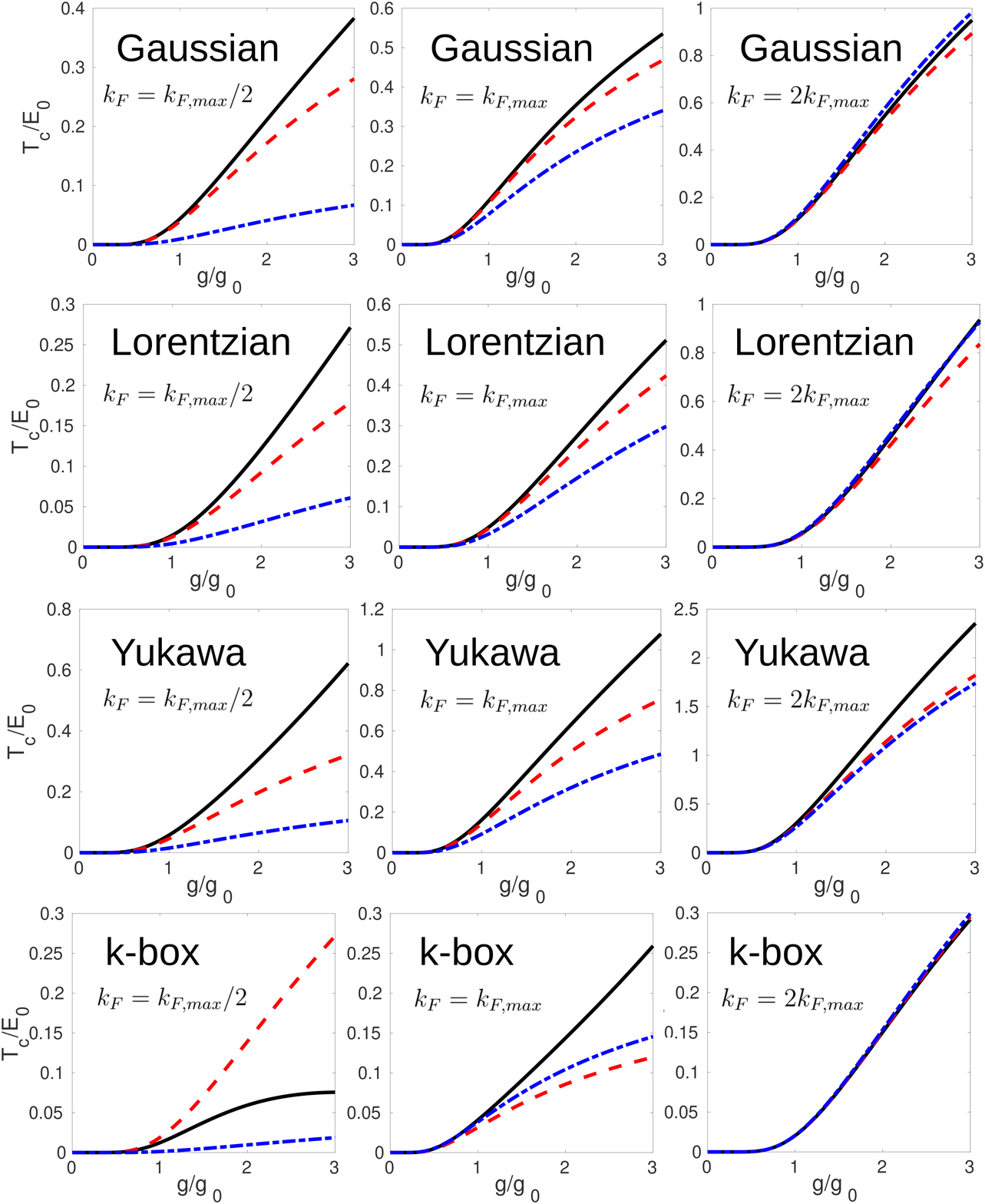}
\caption{Plots of the critical temperature $T_c$ as a function of $g$ for the finite-range potential examples given in Table~I. For each potential $T_c$ has been evaluated at three different doping levels: $k_F=k_{F,max}/2$ (left column); $k_F=k_{F,max}$ (center column); and $k_F=2k_{F,max}$ (right column). In each case we present: (blue/dashed-dotted) $T_c^{(-1)}$, computed by truncating Eq.~(6) at the $1/\lambda$ order; (red/dashed) $T_c^{(0)}$, computed by truncating Eq.~(6) at the $a_0$ order; and (black/solid) $T_c^{(1)}$ in which the 1-st order correction $a_1\lambda$ is included. All energies are reported in units of $E_0=1/2m^*\ell^2$ where $\ell$ is the interaction range, while the coupling constant $g$ is reported in units of $g_0=(2\pi)^2\ell/m^*$. Notice that, in each case, all three curves agree quite well for $g/g_0<0.5$, while for large $g$ the corrections proportional to $\lambda$ become important, especially for $k_F=k_{F,max}/2$ and $k_F=k_{F,max}$. Interestingly, we see that for the case of high doping ($k_F=2k_{F,max}$) the $a_0$ and $a_1$ corrections become significantly less important, consistent with the fact that $\lambda$ decreases for doping levels higher than the superconducting dome. In each of these plots it is possible to convert the horizontal axis from $g/g_0$ to $\lambda$, using the relation $\lambda(k_F)=\tilde{\lambda}(k_F) g/g_0$ and the corresponding numerical values for $\tilde{\lambda}(k_F)$ found in Table.~\ref{table:examples2}.
}
\end{center}
\label{fig:tc_vs_g}
\end{figure}

In Figure~\ref{fig:tc_vs_g}, we plot $T_c$ as a function of $g$ for the finite-range potential examples given in Table~I for three different doping levels: $k_F=k_{F,max}/2$; $k_F=k_{F,max}$; and $k_F=2k_{F,max}$. Our results make clear that the 0-th order approximation, $T_c^{(0)}$, is quite accurate for values of $g/g_0$ up to $0.5$, the value used in Figure~1. Additionally, we see that for the high doping case all three approximations agree remarkably well for the entire range of $g$ considered, consistent with the fact that $\lambda$ decreases for doping values above the superconducting dome. 
It is also interesting to note that, for low-doping, the $a_0$-correction becomes more important for the precise determination of $T_c$, as observed on the left-hand side of the superconducting domes shown in Fig. 1 of the main text.

\end{document}